\documentstyle[12pt,aaspp4]{article}
\begin{document}
\parindent=1.0cm

\title{The Peak Brightness and Spatial Distribution of AGB Stars Near the 
Nucleus of M32}

\author{T. J. Davidge \altaffilmark{1}}

\affil{Canadian Gemini Office, Herzberg Institute of Astrophysics, 
\\ National Research Council of Canada, 5071 W. Saanich Road, \\ Victoria, 
British Columbia, Canada V9E 2E7 \\ {\it email:tim.davidge@hia.nrc.ca}}

\author{F. Rigaut, M. Chun}

\affil{Gemini Observatory, 670 North A'ohoku Place, Hilo, Hawaii, USA 
\\ 96720--2700 {\it email:frigaut@gemini.edu, mchun@gemini.edu}}

\author{ W. Brandner \altaffilmark{1}, D. Potter \altaffilmark{1}, 
M. Northcott, J. E. Graves}

\affil{Institute of Astronomy, University of Hawaii, 2680 Woodlawn Drive, 
\\ Honolulu Hawaii, USA 96822 {\it email:brandner@ifa.hawaii.edu, 
potter@ifa.hawaii.edu, north@ifa.hawaii.edu, graves@ifa.hawaii.edu}}

\altaffiltext{1}{Visiting Astronomer, Gemini Observatory, which is operated 
under contract from the National Science Foundation by AURA Inc. on behalf of 
the Gemini partnership, which consists of the National Science Foundation 
(United States), the Particle Physics and Astronomy Research Council (United 
Kingdom), the National Research Council of Canada, CONICYT (Chile), the 
Australian Research Council, CNPq (Brazil), and CONICET (Argentina).}

\begin{abstract}

	The bright stellar content near the center of the Local Group 
elliptical galaxy M32 is investigated with 0.12 arcsec FWHM $H$ and $K$ images 
obtained with the Gemini Mauna Kea telescope. Stars with $K = 15.5$, which are 
likely evolving near the tip of the asymptotic giant branch (AGB), are resolved 
to within 2 arcsec of the nucleus, and it is concluded that the peak stellar 
brightness near the center of M32 is similar to that in the outer regions of 
the galaxy. Moreover, the projected density of bright AGB stars follows the 
visible light profile to within 2 arcsec of the nucleus, 
indicating that the brightest stars are well mixed throughout the galaxy. 
Thus, there is no evidence for an age gradient, and the 
radial variations in spectroscopic indices and ultraviolet colors that 
have been detected previously must be due to metallicity and/or some other 
parameter. We suggest that either the bright AGB stars formed as part of 
a highly uniform and coherent galaxy-wide episode of star formation, 
or they originated in a separate system that merged with M32.

\end{abstract}


\section{INTRODUCTION}

	As the closest galaxy with characteristics reminiscent of massive 
classical ellipticals (e.g. Wirth \& Gallagher 1984, Kormendy 1985), M32 
provides an unprecedented laboratory for probing the stellar contents of 
early-type galaxies, and testing predictions drawn from integrated spectra. The 
star-forming history of M32 is a matter of on-going debate, although the bulk 
of evidence seems to indicate that an intermediate-age population is present. 
The spectrum of the central few arcsec of M32 can not be represented by that of 
an old star cluster population (O'Connell 1980; Rose 1984; Davidge 1990; Bica, 
Alloin, \& Schmidt 1990) and, while Cole et al. (1998) argue that the 
integrated UV -- visible color of M32 is consistent with an old population, 
they also stress the considerable uncertainties present in the models. The 
spectrum of M32 shows deep H$\beta$ absorption (Burstein et al. 1984), and the 
origin of this feature is of critical importance for establishing if an 
intermediate-age component is present. Brown et al. (2000) resolved hot 
horizontal branch (HB) stars near the center of M32 and, after finding 
that these objects can not on their own explain the strength of H$\beta$ 
absorption, conclude that it will be difficult to explain this feature 
without resorting to a young or a large blue straggler population. 
Finally, Guarnieri, Renzini, \& Ortolani (1997) investigated the stellar 
content of M32 using the metal-rich globular cluster NGC 6553 as a benchmark, 
and found that the brightest infrared source in this cluster is a 
long-period variable that, if viewed at the distance of M32, is more than 0.5 
mag fainter in $K$ than the brightest stars in M32 (Davidge 2000). 

	With a few exceptions (Brown et al. 2000; Davidge 2000), crowding has 
restricted studies of resolved stars to the outer 
regions of M32 (Freedman 1989, 1992; Elston \& Silva 1992; Davidge \& Jones 
1992; Grillmair et al. 1996), and population 
gradients (Cohen 1979; Davidge 1991; O'Connell et al. 1992; Hardy et al. 1994; 
Brown et al. 1998; Ohl et al. 1998) complicate efforts to tie together 
predictions made from resolved stars and integrated spectra. In fact, the 
ultraviolet-visible color of M32 changes by 1.5 mag in the central 10 arcsec 
(Ohl et al. 1998), thus demonstrating the importance of pushing studies of the 
resolved stellar content to smaller and smaller radii. In 
the current paper, images with 0.12 arcsec FWHM are used to resolve bright AGB 
stars to within 2 arcsec of the galaxy center; for comparison, 
the majority of AGB-tip stars resolved by Davidge (2000) are 
at distances in excess of 10 arcsec from the nucleus.

\section{OBSERVATIONS \& REDUCTIONS}

	M32 was observed during the night of 2000 July 5 UT with the University 
of Hawaii adaptive optics (AO) system Hokupa'a, which was mounted at the f/16 
focus of the 8 meter Gemini Mauna Kea telescope. The data were obtained during 
engineering time as part of a program to assess the feasibility of using galaxy 
nuclei as reference beacons for AO compensation. Hokupa'a contains a 36 element 
curvature wavefront sensor and bimorph mirror, and additional details of this 
instrument can be found in Graves et al. (1998). The images were 
recorded with the University of Hawaii QUIRC camera, which contains a $1024 
\times 1024$ Hg:Cd:Te array with an angular scale of 0.02 arcsec pixel$^{-1}$ 
when mounted on Hokupa'a.

	The nucleus of M32 was positioned near the center of the science field 
and was used as the reference source for AO compensation. 
Four 30 sec integrations were recorded in $H$ and $K$ at each of four dither 
positions. The seeing conditions were average for Mauna Kea, 
and the delivered image quality was 0.12 arcsec FWHM in each filter.

	The data were reduced using the procedures described by Davidge \& 
Courteau (1999), and the final $K$ image is shown in Figure 1. The processed 
images were smoothed with a $2 \times 2$ arcsec median filter to produce a 
template of the unresolved body of the galaxy, and this was subtracted from the 
data prior to making photometric measurements. Stellar brightnesses were then 
measured with the PSF-fitting program ALLSTAR (Stetson \& Harris 1988), using 
PSFs and star lists constructed with tasks in DAOPHOT (Stetson 1987). The 
photometric calibration was defined using standard stars from Hunt et al. 
(1998). The outer wings of the PSF extend over a few arcsec, and contain a 
significant fraction ($\sim 10\%$ when $r > 0.5$ 
arcsec) of the total PSF energy. The faint outer portions 
of the PSFs can not be traced in crowded fields, and so the photometric 
measurements were adjusted for this component using corrections obtained from 
AO-compensated images of moderately bright stars with FWHMs comparable to the 
M32 data.

\section{RESULTS}

	The $(K, H-K)$ CMDs of sources in four radial intervals, centered 
on the nucleus of M32, are shown in Figure 2. The error bars show the 
uncertainties predicted by artificial star experiments, and these 
indicate that the dispersion in $H-K$ 
is dominated by photometric errors. These experiments also 
indicate that the trend of bluer $H-K$ colors towards 
fainter brightnesses is a systematic effect resulting from crowding. 

	If two or more stars fall in the same resolution element then they will 
appear as a single bright object, and this is a major concern for 
photometric studies in dense environments. One signature of blending is a 
trend of increasing peak stellar brightness towards progressively smaller 
radii. The region within 2 arcsec of the nucleus contains sources 
that are significantly brighter than those at larger radii, suggesting that 
these objects are artifacts of crowding, likely involving more than two stars. 
However, when $r > 2$ arcsec the peak stellar brightness remains fixed near $K 
= 15.5$ mag, which is in excellent agreement with what is seen in the outer 
regions of the galaxy (Davidge 2000; Freedman 1992; Elston \& Silva 1992). 
Hence, (1) when $r > 2$ arcsec the brightest sources are individual stars, 
and (2) if M32 contains a population that is younger than the brightest 
stars in the main body of the galaxy then it is restricted to the central 2 
arcsec.

	The photometric measurements in Figure 2 are consistent with those 
made by Davidge (2000). Not only is there good agreement between the 
peak stellar brightnesses, but the brightest stars in Figure 2 have $H-K$ 
colors that are similar to the brightest stars in the 
outer field studied by Davidge (2000). Photometric errors and sample 
incompleteness become very large near the RGB-tip, which occurs at $K \sim 
17.8$ mag (Davidge 2000), in the current data and these blur this feature. 
The majority of sources with $K \geq 18$ in Figure 2 are undoubtedly 
blended objects.

	The spatial distribution of bright AGB stars follows the integrated 
light profile of M32 into the central regions of the galaxy. This is 
demonstrated in Figure 3, where the $K$ LFs of sources in the current dataset 
are compared with the outer field $K$ LF from Davidge (2000) and AGB-tip 
star counts from the wide-field infrared survey of Elston \& 
Silva (1992). The outer field LF and AGB-tip counts have been scaled 
to match the mean surface brightness in each radial interval based on the 
Kent (1987) light profile, which was obtained from $r$-band images. The 
outer field data provide a reasonable match to the LFs 
above the RGB-tip (i.e. $K \leq 18$ mag) when $r > 4$ arcsec. 
Crowding hinders the detection of stars fainter than $K = 16.5$ mag between 
2 and 4 arcsec from the nucleus; nevertheless, even in this region the 
counts in the two brightest bins predicted from the outer 
field LF are not significantly different from the observations. These data 
thus extend to smaller radii the conclusion reached by Davidge (2000) that the 
bright AGB population follows the integrated light profile of the galaxy.

\section{DISCUSSION}

	High angular resolution near-infrared images have been used to study 
the bright stellar content near the center of the Local Group elliptical galaxy 
M32. Individual bright stars are resolved to within 2 arcsec of the nucleus, 
and it is concluded that, with the possible exception of the central 2 arcsec 
of the galaxy, where stars have not yet been resolved, the 
AGB-tip remains constant at $K = 15.5$ mag throughout the galaxy. 
This result, coupled with the spatial distribution of the AGB component 
(see below), suggests that the peculiar ultraviolet-visible color profile of 
M32 (O'Connell et al. 1992, Ohl et al. 1998) is not due to an age gradient. 
The brightest stars in the infrared evidently do not 
trace the population that dominates in the ultraviolet. 

	AGB-tip stars are resolved close to the nucleus of M32 because they 
are 2.5 mag brighter in $K$ than the RGB-tip, where the number density of stars 
increases dramatically. Moreover, AGB stars are part of a rapid 
evolutionary phase, and the density of stars within one mag of the AGB-tip, 
which is the portion of the LF where stars are most likely to combine to 
produce sources with brightnesses comparable to, or brighter than, the AGB-tip, 
is relatively low. The low density of bright AGB stars can be demonstrated 
using star counts from the outer region of the galaxy. 
Davidge (2000) found 5 stars within one mag of the AGB-tip in a 0.25 arcmin$^2$ 
field 2.3 arcmin from the nucleus and, after scaling according to the Kent 
(1987) r-band surface brightness profile, the predicted densities of bright 
AGB stars near the nucleus are 0.008 (7.3 -- 13.4 arcsec), 0.016 (4 -- 7.3 
arcsec), and 0.038 (2 -- 4 arcsec) stars per 0.06 arcsec radius resolution 
element. The number of blends expected in each radial interval 
is then 0.01, 0.05, and 0.2. The incidence of blending climbs very quickly 
for fainter objects.

	At a fixed age, the brightness of the AGB-tip is expected to 
vary with metallicity. However, metallicity is likely not a major 
consideration when comparing AGB-tip brightnesses throughout M32. 
Long-slit spectroscopic observations indicate that 
$\Delta$Mg$_2 = -0.06$ between the Davidge (2000) outer field and 2 arcsec if 
$\Delta$Mg$_2$/$\Delta$r$=-0.03$ (Hardy et al. 1994), and this corresponds to 
$\Delta$[Fe/H] = 0.3 -- 0.4 dex using Worthey's (1994) solar metallicity 
models. Bertelli et al. (1994) modelled the evolution of stars up to and 
including the AGB. These models predict that M$_{K}^{AGB-tip}$ changes by 
only 0.1 -- 0.2 mag between z = 0.008 and z = 0.020, in the sense of becoming 
brighter towards higher metallicities.

	The spatial distribution of bright AGB stars follows the r-band surface 
brightness profile measured by Kent (1987) to within 2 arcsec of the nucleus, 
indicating that these stars are uniformly mixed throughout the galaxy. 
If the brightest stars are younger than the underlying body of the galaxy, then 
this result could indicate that these objects formed throughout the galaxy 
during a remarkably coherent episode of star formation, which would require 
that the star-forming material cooled in such a way as to 
follow the mass profile of the underlying galaxy. This is evidently not the 
case in most early-type galaxies, which typically harbour an age gradient 
of size $\Delta$log$t_{Gyr}$/$\Delta$logR = 0.1 (Henry \& Worthey 1999). 
Alternatively, a uniformly mixed population of bright stars is consistent with 
a coeval system, which is a model for M32 suggested by del Burgo et al. (2000).

	Many nearby elliptical galaxies have experienced mergers, and so 
it is worth investigating if the bright stars, which presumably 
formed during intermediate epochs, were 
part of another system that was subsequently accreted by M32. 
If the intermediate-age component accounts for 10\% of M32 by mass (Davidge 
2000), then the accreted galaxy would have had an integrated brightness 
M$_B \leq -13.5$ mag, which overlaps with those of Local 
Group dwarf galaxies. If merger-induced mixing did occur in M32 then (1) 
the age of the brightest AGB stars predicted by models indicates that 
any redistribution of stellar content must have occured 
on a timescale of only a few Gyr (Davidge 2000), and (2) 
any mixing would have to act preferentially on the intermediate-age 
population, and not annihilate the metallicity gradient.
Could an interaction between two galaxies of different 
masses redistribute the stellar content of the smaller object, 
while not completely removing population 
gradients in the larger system? Models of non-dissipative 
mergers between equal-mass systems indicate that the progenitor populations 
are not fully mixed by such interactions (White 1980). However, 
the extent of mixing may be very different if the mass ratio differs 
significantly from unity. Indeed, models in which a low-mass galaxy interacts 
with a disk indicate that the stars in the smaller system 
can be redistributed in much less than a 
Hubble time (Helmi \& White 1999), while not destroying the metallicity 
gradient in the progenitor disk (Quinn, Hernquist, \& Fullagar 1993).
Hernquist \& Quinn (1988) modelled the interaction 
between a low mass system and larger spherical galaxy, and found that 
shells may be produced as the smaller object is consumed by 
the larger system. M32 does not contain shells, and structures of this 
nature may have been removed by interactions with M31.

	A problem with the merging model is that the time scale for dynamical 
friction depends on the inverse of system mass (e.g. Binney 
\& Tremaine 1987), and it is not clear that low-mass systems 
can merge in only a few Gyr. A further complication is that if the AGB stars 
formed in a low-mass system, then they would likely be more metal-poor than the 
majority of stars in M32. If this is the case then the intermediate-age 
component would be even younger than estimated by Davidge (2000), who 
assumed a solar metallicity, thereby producing even tighter requirements on 
the mixing time scale.

\vspace{0.3cm}
	The development of Hokupa'a was supported by the NSF. Sincere 
thanks are extended to Sidney van den Berg, Jim Hesser, and Liese van Zee for 
commenting on an earlier draft of this paper. An anonymous referee also 
provided suggestions that improved the paper.

\clearpage

\begin{center}
FIGURE CAPTIONS
\end{center}

\figcaption
[m32figure1.eps]
{The final $K$ image. North is at the top, and East is to the right.
The field covers $20 \times 20$ arcsec, and the image quality is 0.12 arcsec 
FWHM.}

\figcaption
[m32figure2.eps]
{The $(K, H-K)$ CMDs of sources in four radial intervals centered 
on the nucleus of M32. The sources within 2 arcsec of the nucleus are 
likely not individual objects, but blends of two or more stars. However, 
when $r > 2$ arcsec the peak brightness remains roughly constant at $K = 15.5$, 
indicating that the brightest stars in each interval are not blends. This 
constant peak brightness is also consistent with the absolute photometric 
calibration not varying significantly across the field due to PSF variations. 
The error bars show the uncertainties predicted by artificial star experiments, 
and it is evident that the scatter in $H-K$ is dominated by photometric errors.}

\figcaption
[m32figure3.eps]
{The $K$ LFs in three radial intervals centered on the nucleus of M32 
(solid lines), compared with the LF of sources in the outer field of M32 
observed by Davidge (2000), which has been scaled to match the r-band 
surface brightness in each radial interval (dashed lines). The 
cross in each panel shows the density of stars with $K = 15.5 \pm 0.25$ infered 
from the large field areal survey of Elston \& Silva (1992). 
The error bars show the uncertainties due to counting statistics.}
 
\end{document}